\documentclass[12pt]{iopart}

\usepackage{iopams}
\usepackage{graphicx}
\usepackage{dcolumn}
\usepackage{bm}
\usepackage{color}

\newcommand{\fracc}[2]{\frac{\textstyle{#1}}{\textstyle{#2}}}
\newcommand{\p}{\partial}

\begin{document}
\title{Bianchi-I cosmology from causal thermodynamics}

\author{Eduardo Bittencourt, Leandro G. Gomes, Renato Klippert}
\address{Federal University of Itajub\'a, Av. BPS, 1303, Itajub\'a-MG, 37500-903, Brazil}
\eads{\mailto{bittencourt@unifei.edu.br}, \mailto{lggomes@unifei.edu.br}, \mailto{klippert@unifei.edu.br}}

\date{\today}

\begin{abstract}
We investigate diagonal Bianchi-I spacetimes in the presence of viscous fluids by using the shear and the anisotropic pressure components as the basic variables, where the viscosity is driven by the (second-order) causal thermodynamics. A few exact solutions are presented, among which we mention the anisotropic versions of de Sitter/anti-de Sitter geometries as well as an asymptotically isotropic spacetime presenting an effectively constant cosmic acceleration without any cosmological constant. The qualitative analysis of the solutions for barotropic fluids with linear equations of state suggests that the behaviour is quite general.
\end{abstract}

\maketitle

\section{Introduction}

Bianchi's works on the classification of all admissible Lie algebras upon a $3$-dimensional homogeneous manifold \cite{bianchi} have represented an important step forward in the comprehension of the relationship between geometry and algebra. Few decades later, this interplay shed light into the mathematical features of the four-dimensional spacetime with the development of Einstein's theory of gravity. Since the Belinskii-Khalatnikov-Lifshitz (BKL) conjecture \cite{bkl1,bkl2} on the general behaviour of space-times near the cosmological singularity, the Bianchi models have played an important role in physics, particularly in the early cosmology. It was shown that when one goes back in time towards the initial cosmological singularity, the curvature terms dominate upon the matter ones, and the general metric in this regime is a sort of a generalization of Kasner solutions which, in its turn, are special cases of Bianchi type-I spaces.

Thenceforth, the behaviour of Bianchi type-I cosmological models has been systematically studied for a large class of fluids, assuming non-linear equations of state \cite{rendall}, in the presence of magnetic fields \cite{leblanc}, scalar fields \cite{kitada}, nonlinear Stokesian fluids \cite{dolival} and achieving generic qualitative results \cite{wainwright,calogero}. However, when dissipative terms are taken into account, algebraic relations between the energy-momentum tensor and the geometry are usually considered. This assumption leads to a simplified system of equations where the qualitative behaviour of the solutions can be comprehensively understood \cite{calogero}. On the other hand, by applying the laws of classical thermodynamics, the dissipative processes do not evolve according to hyperbolic equations and, therefore, they are not relativistic and may violate causality.

With this in mind, we shall investigate in this paper the role played by the causal thermodynamics in the description of the dissipative terms of the energy momentum tensor of the cosmic fluid in a Bianchi-I scenario, based on the Israel and Stewart \cite{thermo1,thermo2,thermo3} formulation of the relativistic thermodynamics of irreversible processes, which provides a finite speed of propagation for the viscosity and the heat flow. A somehow truncated version of such thermodynamic theory has already been applied to study the Bianchi type-I cosmologies \cite{bnk,romano1,romano2,coley95,maartens}, while its exact formulation in such context was still missing (up to our knowledge).

The main proposal here concerns the use of such second-order thermodynamical formalism applied to the next-to-the-simplest FLRW class of geometries \cite{haw_ellis} suitable for use as cosmological models, namely the diagonal Bianchi type-I spacetimes. The introduction of a new set of physical variables leads to a more analytically manageable form of the resulting dynamical system, in comparison with previous works on the subject, thus allowing us to obtain new analytical solutions. Additionally, the degrees of freedom of such system can be reduced by making use of first integrals only.

The paper is summarized as follows. In section \ref{bianc-visc}, we present a special set of variables to describe the Einstein equations for a diagonal Bianchi type-I metric in the presence of viscous fluids. These variables have clear mathematical and physical interpretations in terms of standard deviations and kinematic quantities. Section \ref{analytic} provides some analytic solutions with interesting cosmological implications, particularly, for late times. Since the resulting equations are generically non-integrable, in section \ref{qualit} we apply some basic tools of the qualitative theory of differential equations to describe the topological behaviour of the solutions close to the fixed points, including the asymptotic ones, by assuming barotropic fluids as its gravitational sources. In the appendix, we discuss the diagonalizability of Bianchi type-I metrics and present an example which cannot be put into a diagonal form. For the sake of comparison with other references on the subject, the notation adopted here agrees with the one used in \cite{haw_ellis}. We also use geometric units $c=\kappa=1$, unless otherwise stated, where $c$ is the light speed in empty space and $\kappa$ is the Einstein gravitational constant.

\section{Diagonal Bianchi-I with viscous fluids}\label{bianc-visc}
The diagonal Bianchi type-I spacetime is such that the hypersurfaces at constant time describe Euclidean spaces. Its infinitesimal line element is commonly expressed in Cartesian coordinates as
\begin{equation}
\label{ds2-bian_l}
{\rm d}s^2=-{\rm d}t^2+[l_{1}(t)]^2{\rm d}x^2+[l_{2}(t)]^2{\rm d}y^2+[l_{3}(t)]^2{\rm d}z^2,
\end{equation}
where $l_i=l_i(t)$, $i=1,2,3$, are the scale factors in each spatial direction. When $t={\rm const.}$, the $3$-volume of these hypersurfaces suggests the definition of a mean scale factor provided by the geometric mean of the $l_i$'s:
\begin{equation}
\label{a_li}
a(t)\doteq \sqrt[3]{l_1\,l_2\,l_3}.
\end{equation}
From this, we compute the difference between the expansion rate in the $x^i$-direction and the mean expansion rate:
\begin{equation}
\label{sigma_a_li}
\sigma_i\doteq\frac{\dot l_i}{l_i}-\frac{\dot a}{a},
\end{equation}
where the dot means time derivative, as $\dot X \doteq {\rm d}X/{\rm d}t$. These $\sigma_i$'s measure how the scale factor along each axis deviates from its mean value. Throughout the text we shall refer to $\dot a/a$ as the Hubble expansion rate (or the Hubble parameter) in analogy to the isotropic FLRW models, since it is the mean value of the expansion rates in different spatial directions, that is
\begin{equation}
\label{h_a_li}
H\doteq\frac{\dot a}{a}=\frac{1}{3}\left(\frac{\dot l_1}{l_1}+\frac{\dot l_2}{l_2}+\frac{\dot l_3}{l_3}\right).
\end{equation}
We can then cast the line element in terms of $a$ and  $\sigma_i$, as
\begin{equation}
\label{ds2-bian}
{\rm d}s^2=-{\rm d}t^2+a^2(t)\left[{\rm e}^{2\int\sigma_1(t){\rm d}t}{\rm d}x^2+{\rm e}^{2\int\sigma_2(t){\rm d}t}{\rm d}y^2+{\rm e}^{2\int\sigma_3(t){\rm d}t}{\rm d}z^2\right].
\end{equation}
From equations (\ref{sigma_a_li}) and (\ref{h_a_li}), it follows that the $\sigma_i$'s must satisfy the constraint
\begin{equation}
\label{con_dhi}
\sigma_1+\sigma_2+\sigma_3=0.
\end{equation}
For a time-like vector field of co-moving observers (namely, $V^{\mu}=\delta^{\mu}_0$), the $\sigma_i$'s coincide with the eigenvalues of the mixed form $\sigma^{\mu}{}_{\nu}$ of the symmetric traceless shear tensor $\sigma_{\mu\nu} \doteq \frac{1}{2}(\delta^{\gamma}_{\mu} + V^{\gamma}V_{\mu} )(\delta^{\rho}_{\nu} + V^{\rho}V_{\nu})( V_{\gamma;\rho} + V_{\rho;\gamma} ) - \frac{1}{3} V^{\alpha}{}_{;\alpha} ( g_{\mu\nu} + V_{\mu} V_{\nu} )$ for the spacelike eigenvectors.

The standard deviation associated to the Hubble parameter is a measure of the difference between the Bianchi-I model and its plane and isotropic counterpart ($\sigma_i = 0$). The variable standing for this geometrical anisotropy is expressed as
\begin{equation}
\label{sigma}
\sigma\doteq\sqrt{\sum_i\left(\frac{\dot l_i}{l_i}-H\right)^2}=\sqrt{\sigma_1^2+\sigma_2^2+\sigma_3^2}.
\end{equation}

\begin{figure}
\begin{center}
\includegraphics[width=7cm,height=7cm]{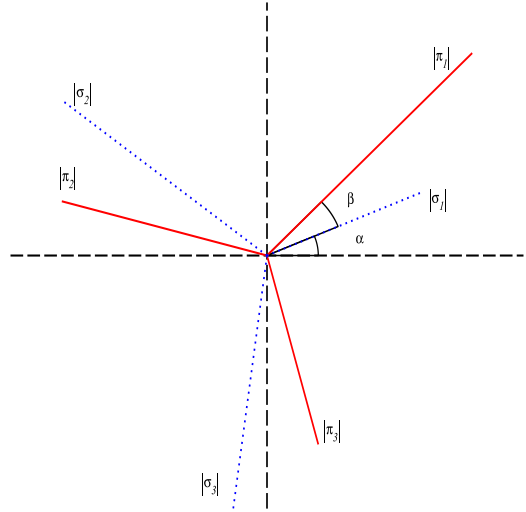}
\end{center}
\caption{Colors online. Comparative analysis of the principal directions of the metric anisotropies (dotted-blue)
and those of the anisotropic pressure (solid-red). Each component of the metric anisotropy $\sigma_i$ is separated by an angle $2\pi/3$, see equations (\ref{par-dhi-alpha}). The same happens with the components $\pi_i$ of the anisotropic pressure [see equations (\ref{par-dpi-beta})].}
\label{fig1}
\end{figure}

With the help of a time dependent angular coordinate $\alpha=\alpha(t)$, see\ figure \ref{fig1}, we can parameterize the plane described by equation (\ref{con_dhi}) as
\begin{equation}
\label{par-dhi-alpha}
\left\{\begin{array}{lcl}
\sigma_1&=&\sqrt{\frac{2}{3}}\,\sigma\sin\alpha\,,\\[2ex]
\sigma_2&=&\sqrt{\frac{2}{3}}\,\sigma\sin\left(\alpha+\frac{2\pi}{3}\right),\\[2ex]
\sigma_3&=&\sqrt{\frac{2}{3}}\,\sigma\sin\left(\alpha+\frac{4\pi}{3}\right).
\end{array}\right.
\end{equation}

The homogeneity of the Bianchi type-I geometry described above demands the matter content to be independent of the spatial coordinates. Since the Einstein tensor for the Bianchi-I metric (\ref{ds2-bian}) is diagonal, the compatibility with the Einstein's equations implies a diagonal form for the energy momentum tensor, with respect to coordinates which are adapted to the comoving vector $V^{\mu}=\delta^{\mu}_0$. Then, we consider

$$T^{\mu}{}_{\nu}={\rm diag}(-\rho, p+\pi_1, p+\pi_2, p+\pi_3),$$
with $\rho=\rho(t)$ as the energy density, $p=p(t)$ as the total isotropic pressure\footnote{The total pressure $p$ may include both the thermodynamic $P$ and the bulk ${\cal P}$ pressures. However, as suggested by \cite{coley95} for a particular case, we shall neglect any possible contribution from ${\cal P}$.} and $\pi_i=\pi_i(t)$, $i=1,2,3$, as the eigenvalues of the symmetric and traceless anisotropic pressure tensor $\pi_{\mu\nu}$ for the space-like eigenvectors. Thus,
\begin{equation}
\label{con_dpi}
\pi_1+\pi_2+\pi_3=0.
\end{equation}

We now have the standard deviation of the pressures in different directions as a measure of the magnitude of the anisotropy in the matter content
$$\Pi\doteq \sqrt{\pi_1^2+\pi_2^2+\pi_3^2},$$
and a similar parametrization of the plane defined by equation (\ref{con_dpi}) can be performed
[see also figure \ref{fig1}]
\begin{equation}
\label{par-dpi-beta}
\left\{\begin{array}{lcl}
\pi_1&=&\sqrt{\frac{2}{3}}\,\Pi\,\sin(\alpha+\beta)\,,\\[2ex]
\pi_2&=&\sqrt{\frac{2}{3}}\,\Pi\,\sin\left(\alpha+\beta+\frac{2\pi}{3}\right),\\[2ex]
\pi_3&=&\sqrt{\frac{2}{3}}\,\Pi\,\sin\left(\alpha+\beta+\frac{4\pi}{3}\right),
\end{array}\right.
\end{equation}
where $\beta=\beta(t)$, which we name the {\it anisotropy phase}, is the angle between the Euclidean vectors
$\vec\Pi=(\pi_1,\pi_2, \pi_3)$ and $\vec \sigma=(\sigma_1,\sigma_2, \sigma_3)$. Thus,
$$\cos\beta=\frac{\vec\sigma\cdot\vec\Pi}{\sigma\,\Pi}.$$
The anisotropy phase $\beta$ measures the coupling between the geometric and the matter anisotropies.

The Einstein equations with a cosmological constant $G^{\mu}{}_{\nu} = T^{\mu}{}_{\nu} + \Lambda\delta^{\mu}{}_{\nu}$ lead, through the same procedure as in the FLRW case, to the following system of equations:
\begin{eqnarray}
&&3H^2=\rho+\frac{\sigma^2}{2}+\Lambda,\label{ee1}\\[2ex]
&&\dot\rho + 3H(\rho+p)+\vec\sigma\cdot\vec\Pi=0,\label{ee2}\\[2ex]
&&\dot\sigma_i+3H\sigma_i=\pi_i.\label{ee3}
\end{eqnarray}
By using formulas (\ref{par-dhi-alpha}) and (\ref{par-dpi-beta}), it is possible to rewrite the three implicit equations encoded in (\ref{ee3}) in terms of two other ones:
\begin{eqnarray}
&&\dot{\sigma}+3H\sigma=\Pi\cos\beta,\label{ee31}\\[1ex]
&&\sigma\dot\alpha=\Pi\sin\beta.\label{ee32}
\end{eqnarray}
Equation (\ref{ee1}) is a modified version of the Friedmann's isotropic one, where the difference is the geometric anisotropy term $\sigma^2/2$ contribution to the total energy density of the universe. The continuity equation is represented by equation (\ref{ee2}). Note that, with this set of variables, we end up with first order differential equations only.

It should also be remarked that equations (\ref{ee1})--(\ref{ee3}) correspond to an open system. As it was previously mentioned, there are several ways to fix the arbitrariness present in these equations, which are usually performed by taking into account algebraic relations between the metric and the fluid components (see \cite{calogero} and references therein). We shall use here the causal thermodynamics \cite{thermo1,thermo2,thermo3} to play this role in what concerns the dissipative terms, by adding differential equations instead of algebraic ones.

The dynamical equation for $\pi_i$ is given in \cite{thermo1} as
\begin{equation}
\label{therm_pi}
\tau\dot\pi_i+\pi_i=-\xi\sigma_i,
\end{equation}
where the thermodynamic parameters $\tau$ and $\xi$, phenomenologically dependent upon the thermodynamic quantities, are the relaxation time and the shear viscosity, respectively. In terms of the standard deviations $\Pi$ and $\sigma$, and the anisotropy phase $\beta$, we can rewrite equations (\ref{therm_pi}), as
\begin{eqnarray}
&&\tau\dot{\Pi}+\Pi=-\xi\sigma \cos\beta,\label{stand_pi}\\[2ex]
&&\dot\beta+\left(\frac{\Pi}{\sigma}-\frac{\xi}{\tau}\frac{\sigma}{\Pi}\right)\sin\beta=0.\label{ev_beta}
\end{eqnarray}
From the combination of equations (\ref{ee31}), (\ref{stand_pi}) and (\ref{ev_beta}), we can find a first integral of the system of differential equations from the ``norm of the vector product'' between $\vec\Pi$ and $\vec\sigma$
\begin{equation}
\label{const_0}
\Pi \, \sigma\,\sin\beta=\fracc{L_0\,{\rm e}^{-\int{\frac{{\rm d}t}{\tau(t)}}}}{a^3},
\end{equation}
where $L_0$ is an integration constant. From equations (\ref{stand_pi}) and (\ref{ev_beta}) we also see that, when the system approaches a thermodynamical equilibrium configuration in an expanding universe scenario, we have that $\beta$ goes to $0$ or $\pi$ if the mean values of the anisotropies are both nonzero.

There is another interesting relation between the anisotropies provided by the above equations which is

\begin{equation}
\label{lean_rel}
\frac{{\rm d}}{{\rm d}t}\left(\Pi^2+\frac{\xi}{\tau}\sigma^2\right)=-\frac{2}{\tau}\left(\Pi^2+3H\xi\sigma^2\right).
\end{equation}
For $\xi,H\geq0$ and $\tau>0$, this means that the term within the parenthesis on the left-hand side has an upper bound and, therefore, the anisotropies cannot grow indefinitely.

Differently from previous works on this subject, which consider a phenomenological relation between the thermodynamic parameters $\tau$ and $\xi$ and the energy density $\rho$ (see \cite{bnk,coley95} for instance), we will assume them as constants for simplicity, unless otherwise stated. This choice will change qualitatively the behaviour of the solutions as we shall see afterwards, since $\tau$ and $\xi$ may contribute even if the energy density is small.

\section{Exact solutions}\label{analytic}
In this section, we concentrate our efforts on the task of finding classes of exact solutions of physical interest for the system of equations (\ref{ee1})--(\ref{ev_beta})\footnote{In fact, there are two equivalent linearly independent subsets of these equations which describe the whole system at its regular configurations. They are given by equations (\ref{ee1})--(\ref{ee3}) and (\ref{therm_pi}), or, equivalently, by the system (\ref{ee1}), (\ref{ee2}), (\ref{ee31}), (\ref{ee32}), (\ref{stand_pi}) and (\ref{ev_beta}). In the cases for which this latter system turns out to be singular, one must return to the former one, as we shall see later.}.

\subsection{Bianchi-I with perfect fluids revisited}\label{bianc-perf}
As a matter of comparison with our approach through the standard deviation variables $\sigma$ and $\Pi$, we first revisit the equations for Bianchi-I models with perfect fluids, see for instance \cite{wainwright} or section 14.4 of \cite{ExactSolutions} for more details. Then, for $\tau$ and $\xi$ both identically equal to zero, the aforementioned system of equations can be completely integrated, resulting in
\begin{equation}
\label{sig_kasn}
\sigma=\sqrt{2}\,\frac{\Omega_A}{a^3},\quad\mbox{and}\quad \alpha=\alpha_0,
\end{equation}
where $\Omega_A$ and $\alpha_0$ are integration constants; from equation (\ref{ee1}), we may interpret $\Omega_A$ as the density parameter associated to the contribution of the anisotropy to the total energy of the universe.

In the absence of matter, i.e. $p=\rho=0$, the Kasner solutions should be recovered. Using equation (\ref{sig_kasn}), the integration of the modified Friedmann equation yields
\begin{equation}
\label{kasn_a}
a(t)=\left\{
\begin{array}{ll}
\left(\frac{\Omega_A}{\Lambda}\right)^{\frac{1}{6}}\,\sinh^{\frac{1}{3}}[\pm\sqrt{3\Lambda}(t-t_0)+C],&\quad\mbox{for}\quad \Lambda>0,\\[1ex]
\pm[\sqrt{3\Omega_A}(t-t_0)+C]^{\frac{1}{3}},&\quad\mbox{for}\quad \Lambda=0,\\[1ex]
\left(\frac{\Omega_A}{|\Lambda|}\right)^{\frac{1}{6}}\,\sin^{\frac{1}{3}}[\pm\sqrt{3|\Lambda|}(t-t_0)+C],&\quad\mbox{for}\quad \Lambda<0,
\end{array}\right.
\end{equation}
where $C$ is an integration constant. The positive sign holds for $H>0$ and the negative sign for $H<0$. Therefore, the directional scale factors are
\begin{equation}
\label{kasn_li}
l_i = l_{i,0} \left(\frac{a}{a_0}\right)^{3\zeta_i} \left( \frac{\sqrt{\Omega_A} + \sqrt{\Omega_A + \Lambda a^6}}{\sqrt{\Omega_A} + \sqrt{\Omega_A + \Lambda a_{0}^6}} \right)^{\frac{1}{3}-\zeta_i},
\end{equation}
where $l_{i,0}$ are constant, $a_0\doteq a(t=t_0)$ and $\zeta_i\doteq2\sin(\alpha_i)$ with $\alpha_i=\alpha_0 +2(i-1)\pi/3$ and $i=1,2,3$. It is direct to show that $\zeta_1+\zeta_2+\zeta_3=1$ and $\zeta_1^2 + \zeta_2^2 +\zeta_3^2=1$. When $\Lambda=0$, this is indeed the Kasner solution (see reference \cite{bkl1}).

When the matter content is present and we take the pressure as a known function of $a$ and $\rho$, i.e. $p=p(a,\rho)$, then we can use the continuity equation (\ref{ee2}) to write the energy density in terms of the mean scale factor
\begin{equation}
\label{cont_rho_a}
\frac{{\rm d}\rho}{{\rm d}a}=-\frac{3}{a}[\rho+p(a,\rho)].
\end{equation}
Thus, the modified Friedmann equation can be used to obtain an implicit expression for $a(t)$:
\begin{equation}
\label{imp_a_kasn_fl}
\sqrt{3}\int{\frac{{\rm d}a}{a\sqrt{\Lambda +\frac{\Omega_A}{a^6} + \rho(a)}}}=t.
\end{equation}
From equation (\ref{par-dhi-alpha}), we compute the integral of the $\sigma_i$'s as follows
\begin{equation}
\label{int_sigma_kasn_fl}
\int\sigma_i\,{\rm d}t=\sqrt{\frac{2}{3}}\sin\left(\alpha_i\right)\int\sigma {\rm d}t=2\sin\left(\alpha_i\right)\int\frac{{\rm d}a}{a\sqrt{1+\bar\Lambda a^6+\Upsilon(a^3)}},
\end{equation}
where $\bar\Lambda\doteq\Lambda/\Omega_A$ and $\Upsilon(u)\doteq u^2\rho(u^{\frac{1}{3}})/\Omega_A$. Therefore, the line element becomes in this case
\begin{equation}
\label{line_el_kasn_fl}
{\rm d}s^2=-\left[\frac{3a^4}{\Omega_A+(\Lambda+\rho)a^6}\right]{\rm d}a^2+a^2\sum_{i=1}^3{\rm e}^{\frac{4}{3}\sin\left(\alpha_i\right)G(a^3)}\, ({\rm d}x^i)^2,
\end{equation}
where $G(u)=\int\left[u\sqrt{1+\bar\Lambda u^2+\Upsilon(u)}\right]^{-1}{\rm d}u$.

\subsection{de Sitter/anti-de Sitter scale-factor}\label{desitter}
Let us suppose the mean scale-factor behaves as \(a(t)=a_o{\rm e}^{Ht}\) with \(H\) constant, recalling that the thermodynamic parameters are constants as well. We also make use of two pairs of new variables: \(x_i={\rm e}^{t/\tau}\pi_i\), and \(\,y_i=a^3\sigma_i\) with \(i=1,\,2\). It then follows that the dynamical equations (\ref{ee3}) and (\ref{therm_pi}) are given in the form
\begin{equation}
\left\{\begin{array}{l}
\dot x_i=-(\xi/\tau)\,{\rm e}^{t/\tau}a^{-3}y_i,\\
\dot y_i={\rm e}^{-t/\tau}a^3x_i.
\end{array}\right.
\end{equation}
By differentiating \(y_i\) twice, the above dynamics immediately yields the equations
\begin{equation}
\label{dyn_yi}
\ddot y_i+\left(\frac1\tau-3H\right)\dot y_i+\frac{\xi}{\tau}\,y_i=0
\end{equation}
of two identical decoupled damped oscillators, where the ``free frequency'' is \(\sqrt{\xi/\tau}\) and the ``friction parameter'' is \((1/2)(1/\tau-3H)\). These equations are therefore solved according with the sign of \(\mu=(1/4)(1/\tau-3H)^2-\xi/\tau\). It then follows, for $\mu>0$, that
\begin{equation}
\label{sol_ren_k_l_0}
\fl
\left\{\begin{array}{lcl}
\pi_i&=&\frac12{\rm e}^{-\frac12\left(3H+\frac1\tau\right)t}\,\left[A_i\left(3H-\frac1\tau+2\sqrt \mu\right){\rm e}^{\sqrt \mu\,t}+B_i\left(3H-\frac1\tau-2\sqrt \mu\right){\rm e}^{-\sqrt \mu\,t}\right],\\[2ex]
\sigma_i&=&{\rm e}^{-\frac12\left(3H+\frac1\tau\right)t}\,\left[A_i{\rm e}^{\sqrt \mu\,t}+B_i{\rm e}^{-\sqrt \mu\,t}\right],
\end{array}\right.
\end{equation}
while for $\mu<0$, we have
\begin{equation}
\label{sol_ren_k_e_0}
\fl
\left\{\begin{array}{lcl}
\pi_i&=&\frac12{\rm e}^{-\frac12\left(3H+\frac1\tau\right)t}\,\left[\left(3H-\frac1\tau\right)A_i\,\cos\,\sqrt{|\mu|}\,t + 2\sqrt{|\mu|}B_i\,\sin\,\sqrt{|\mu|}\,t\right],\nonumber\\[2ex]
\sigma_i&=&{\rm e}^{-\frac12\left(3H+\frac1\tau\right)t}\,\left[A_i\,\cos\,\sqrt{|\mu|}\,t+B_i\,\sin\,\sqrt{|\mu|}\,t\right],
\end{array}\right.
\end{equation}
and for $\mu=0$, we obtain
\begin{equation}
\label{sol_ren_k_s_0}
\fl
\left\{\begin{array}{lcl}
\pi_i&=&{\rm e}^{-\frac12\left(3H+\frac1\tau\right)t}\,\left[B_i+\frac12\left(3H-\frac1\tau\right)(A_i+B_i\,t)\right],\nonumber\\
\sigma_i&=&{\rm e}^{-\frac12\left(3H+\frac1\tau\right)t}\,\left[A_i+B_i\,t\right].
\end{array}\right.
\end{equation}

By using the definition of $y_i$, the obtention of the energy density from the Friedman's equation (\ref{ee1}) is direct
\begin{equation}
\label{rho_a_exp}
\rho=3H^2-\Lambda-\frac{\sigma^2}{2}.
\end{equation}
The physical requirement of non-negativity of the energy density \(\rho\ge0\) then imposes a bounded behaviour for the geometrical anisotropy magnitude \(\sigma\), which in turn impose limitations on the time domains for the solutions given from equations (\ref{sol_ren_k_l_0})--(\ref{sol_ren_k_s_0}): if \(3H+1/\tau>0\), then \(t\ge\bar t\), where \(\bar t\) is the time for which \(\rho=0\) from equation (\ref{rho_a_exp}); if \(3H+1/\tau<0\), then \(t\le\bar t\). From inspection of these solutions, we obtain that \(\Pi\) is similarly bounded.

The mean pressure $p$ can be calculated from equation (\ref{rho_a_exp}) with the help of the conservation equation (\ref{ee2}) as
\begin{equation}
\label{p_a_exp}
p=\rho+2\Lambda- 6 H^2=\Lambda- 3 H^2-\frac{\sigma^2}{2}.
\end{equation}
As a consistency check of the method, notice that, for $\sigma=0$, we recover the isotropic de Sitter case with $\Lambda = 3 H^2$.

From equation (\ref{sol_ren_k_e_0}), one should notice that the particular case $3H\tau=-1$ with $\tau\xi>1$ leads to a pure oscillatory behaviour of the anisotropies $\sigma_i$ and $\pi_i$ with frequency $\omega=\sqrt{\xi/\tau}$. Therefore, Bianchi-I models filled in with anisotropic fluids allow for solutions represented by periodic orbits, as already noticed by van den Hoogen and Coley in \cite{coley95}.

\subsection{Phenomenological thermodynamic parameters}
The cases described above present an accelerated scenario which may possibly represent only part of the time evolution of the universe. Let us  comment on how one could produce a more realistic situation from such exact solutions by supposing that the thermodynamic parameters $\tau$ and $\xi$ have a time dependence, since the thermodynamic quantities they originally depend upon are also functions of time. We assume that these parameters satisfy the relation
\begin{equation}
\label{evol_tau}
\dot\tau=(3H+\xi)\tau,
\end{equation}
where the mean scale-factor $a$ is considered to be a known function of time (and thus, $H$ is also known). We introduce variables \((x_i,\,y_i)\) which resemble the ones in the previous section:
\[\left\{\begin{array}{l}
x_i={\rm e}^{\int[1/\tau(t)]{\rm d}t}\pi_i,\\[1ex]
y_i=a^3\sigma_i.
\end{array}\right.\]
The dynamics, in terms of these variables, takes the form
\[\left\{\begin{array}{l}
\dot x_i=-\frac\xi\tau\,{\rm e}^{\int{\rm d}t/\tau}a^{-3}y_i,\\[1ex]
\dot y_i={\rm e}^{-\int{\rm d}t/\tau}a^3x_i.
\end{array}\right.\]
By differentiating \(y_i\) twice, we again obtain that each \(y_i\) satisfies equation (\ref{dyn_yi}), but now the coefficients are not constant anymore. In order to solve this equation, we restate it in terms of new variables \(w_i=f\,y_i\), where the proportionality factor \(f=a^{3/2}\,{\rm e}^{-\frac12\int{\rm d}t/\tau}\) was chosen such that the coefficient of the term \(\dot w_i\) is zero. The resulting equation for \(w_i\) takes the form \(\ddot w_i=[\ddot f-(\xi/\tau)f]w_i\) as
\begin{eqnarray}
\ddot w_i&=&\frac14\left[3(2\dot H+3H^2)+\frac{\textstyle1+2\dot\tau}{\textstyle\tau^2}-\frac{\textstyle4\xi+6H}{\textstyle\tau}\right]w_i\nonumber\\
&=&{\frac14\left[3(2\dot H+3H^2)+\frac{\textstyle1-2\dot\tau}{\textstyle\tau^2}+\frac{\textstyle6H}{\textstyle\tau}\right]w_i},
\label{wi}
\end{eqnarray}
where the last equality follows from the choice \(\xi=\dot\tau/\tau-3H\) in equation (\ref{evol_tau}). It is thus apparent that one of the solutions of equation (\ref{wi}) is \(w^{(1)}=a^{3/2}\,{\rm e}^{\frac12\int{\rm d}t/\tau}\). The other solution is obtained from this one as \(w^{(2)}=u(t)\,w^{(1)}\), from which \(v=\dot u\) satisfies a first order differential equation. We then obtain the general solutions of equation (\ref{wi}) as
\begin{equation}
w_i=A_i\,a^{3/2}\,{\rm e}^{\frac12\int{\rm d}t/\tau}+B_i\,\int{\rm d}t\,a^{-3}\,{\rm e}^{-\int{\rm d}t/\tau},
\label{w3}
\end{equation}
where $A_i,B_i$ are constants, for \(i=1,2\). The mean pressure $p$ is given in this case as
\begin{equation}
p=\rho+2\Lambda-2H^2-\frac{2}{3}\dot H,
\end{equation}
and $\rho$ is obtained in a similar way as in the previous subsection \ref{desitter}.

For the particular $B_i=0$ case, it follows that
\begin{equation}
\label{sigma_phen}
\sigma_i=A_i\,a^{-3}{\rm e}^{\int{\rm d}t/\tau}=\tau\,\pi_i.
\end{equation}
Note that $\sigma$ and $\Pi$ are linearly related, and their ratio is an arbitrary function of time. The viscoelastic proposal $\xi=\kappa_1\rho \tau$ with $\kappa_1$ constant \cite{bnk} can be regarded as a solution of the type of equation (\ref{sigma_phen}) with $\beta=\pi/2$, $3H\tau(2+\gamma) = 2$ and
\begin{equation}
\dot\xi=\xi^2-\frac{2\kappa_1(\gamma-1)}{2+\gamma}\,\rho,
\end{equation}
where $p/\rho+1=\gamma$ (which is not being assumed to be constant), while
\begin{equation}
\sqrt{\kappa_1\rho}\,\sigma^2 a^3 {\rm e}^{\frac{3}{2}\int{(2+\gamma)H\,{\rm d}t}}=L_0.
\end{equation}
In principle, $\tau$ could be chosen such that the solutions presented above could be important in, at least, two distinct periods of the universe history: the inflationary era and the current accelerating expanding phase. The details are left for a forthcoming paper.

\subsection{Cosmological constant from a residual anisotropy}
Another interesting solution of equations (\ref{ee1})--(\ref{ev_beta}) can be obtained if we consider the anisotropic pressure standard deviation to be proportional to the metric anisotropy one, i.e., $\Pi=k_0\sigma$, but now with a constant $k_0$. The angular equations (\ref{ee32}) and (\ref{ev_beta}) can be completely integrated, yielding
\begin{equation}
\tan\frac{\beta}{2}=A_o{\rm e}^{-k_1(t-t_o)},
\end{equation}
\begin{equation}
\alpha=\alpha_o-\frac{k_0}{k_1}(\beta-\beta_o),
\end{equation}
where $A_o,t_o,\alpha_o,\beta_o$ are integration constants and $k_1\doteq k_0-\frac{\xi}{\tau}(k_0)^{-1}$. Equations (\ref{ee31}) and (\ref{const_0}) then give
\begin{equation}
\label{sigma_exp}
\sigma=\frac{\sigma_o}{a^3}{\rm e}^{k_0\int^{t}_{t_o}\cos\beta {\rm d}t},
\end{equation}
\begin{equation}
\label{scale_exp}
a(t)={\rm e}^{k_2(t-t_o)}\left(\frac{1+A_o^2{\rm e}^{-2k_1(t-t_o)}}{1+A_o^2}\right)^{k_3},
\end{equation}
where $\sigma_o$ is an integration constant, $k_2\doteq\frac{1}{3}(2k_0-k_1+\frac{1}{\tau})$ and $k_3\doteq\frac{1}{3}(\frac{2k_0}{k_1}-1)$.  The remarkable result from this solution is the Hubble rate. From equation (\ref{scale_exp}), the ratio $\dot a/a$ can be written down as
\begin{equation}
\label{hub_rat_exp}
H = k_2 - \frac{2A_o^2\, k_1\,k_3}{A_o^2 + {\rm e}^{2k_1(t-t_o)}} = \frac{1}{3\tau} + \frac{1}{3k_0} \left(k_0^2 + \frac{\xi}{\tau} \right) \frac{{\rm e}^{2k_1(t-t_o)}-A_o^2}{{\rm e}^{2k_1(t-t_o)}+A_o^2}.
\end{equation}
If $k_1(t-t_o)\gg1$, we see that $H$ tends to a non-zero constant, which depends only on the thermodynamic parameters of the fluid, thus mimicking the cosmological constant role in the Friedmann equation. Therefore, the apparent behaviour of a cosmic acceleration of our Universe might be the relics of an anisotropic fluid in the past. From the combination of the equations (\ref{sigma_exp})--(\ref{hub_rat_exp}), one can easily read the energy density from the modified Friedmann equation (\ref{ee1}), and thus, the continuity equation (\ref{ee2}) determines the mean pressure.

\section{Outlines of the qualitative analysis for barotropic fluids}\label{qualit}

Qualitative analysis of Bianchi-I systems complemented with the causal thermodynamics equations have been considered at least from the early $80$'s, with the work of  Belinskii, Nikomarov and Khalatnikov \cite{bnk}. There we have the relations $p=(\gamma -1)\rho$, $\xi = a_0 \rho^b$ and $\tau = a_1 \rho^{b-1}$, with $\gamma, a_0, a_1, b$ constants, and an analysis of the system of equations (\ref{ee1})--(\ref{ev_beta}) in the variables $X=-\sigma \Pi \cos \beta$ and $Y=\Pi^2$ were made, with a particular attention to the limit $H,\rho \to+\infty$ and $\ell_i\to0$, with $i=1,2,3$ (the Big Bang singularity).

As an example of further investigations on the subject, van den Hoogen and Coley \cite{coley95} analysed such system as a residual case of Bianchi-V spacetimes. In our representation, this is equivalent to set $\alpha = 0 = \beta$. With the help of the algebraic relations $p=(\gamma -1)\rho$, $b_2 H \xi = 2 a_2 \rho$ and $2 b_2 \tau H=1$, with $\gamma, a_2, b_2$ constants, it is then straightforward to obtain the system (\ref{ee1})--(\ref{ev_beta}) as a planar one, which is given by equations (3.29a) and (3.29b) in \cite{coley95}. In terms of the dimensionless variables $ x = \rho/3 H^2$, $\sqrt{3}\Sigma = \sqrt{2}\sigma/H$ and $z =  \Pi/(6 \sqrt{6}\, H^2)$, then the modified Friedmann equation (\ref{ee1}) reads $4 - 4 x - \Sigma^2 =0$. This model has a remarkable periodic orbit for positive and small values of the parameter $B_8=4 b_2 - 3(3\gamma-2)$.

In order to have a deeper understanding of the system of equations developed in section \ref{bianc-visc}, we shall analyse its qualitative behaviour for the special class of barotropic fluids with a linear equation of state. But instead of considering the thermodynamic parameters as satisfying phenomenological relations with the dynamical variables, we shall take $\xi$ and $\tau$ as constants, with $\tau>0$. This will allow us to determine the eigenvalue problem for the linearised system, the outcome of which can be compared with the asymptotic behaviour of the exact solutions previously discussed.

In terms of the standard deviation variables $\sigma$ and $\Pi$, the angles $\alpha$ and $\beta$, the expansion coefficient $\theta\doteq3H$ and their respective equations of motion, we can define the following dynamical system\footnote{Although the dynamical system is defined in ${\cal M}$, the physically meaningful points $X\in{\cal M}$ are those which belong to the submanifold $\mathbb{R}\times\mathbb{R}_+^2\times\mathbb{T}^2$, since \(\sigma\ge0\) and \(\Pi\ge0\).}
 in ${\cal M}=\mathbb{R}^3\times\mathbb{T}^2$:
\numparts
\begin{eqnarray}
\dot{\theta}&=&-\frac{1}{2}\left[\theta^2+\frac{3}{2}\sigma^2+3(p-\Lambda)\right],\label{sys_dyn_3}\\[1ex]
\dot{\sigma}&=&-\theta\sigma+\Pi\cos\beta,\label{sys_dyn_1}\\[1ex]
\dot{\Pi}&=&-\frac{1}{\tau}\Pi-\frac{\xi}{\tau}\sigma\cos\beta,\label{sys_dyn_2}
\end{eqnarray}
\endnumparts
where the total pressure $p$ is assumed to be a given function of $\theta$ and $\sigma$, and
\numparts
\label{sys_dyn_ang}
\begin{eqnarray}
\dot\beta&=&-\left(\frac{\Pi}{\sigma}-\frac{\xi}{\tau}\frac{\sigma}{\Pi}\right)\sin\beta,\label{sys_dyn_5}\\[1ex]
\dot\alpha&=&\frac{\Pi}{\sigma}\sin\beta.\label{sys_dyn_4}
\end{eqnarray}
\endnumparts
Note that equation (\ref{sys_dyn_3}) is obtained from the substitution of the modified Friedmann equation (\ref{ee1}) in the Raychaudhuri equation. Finally, one can see that $\alpha$ does not appear on the right-hand side of any equation above; therefore, equation (\ref{sys_dyn_4}) can be solved by quadrature once $\sigma,\Pi,\beta$ are known.

The equilibrium points $X_o\in {\cal M}$ are obtained by setting the right-hand side of equations (\ref{sys_dyn_3})--(\ref{sys_dyn_4}) equal to zero, which, for $\sigma_o\neq0$, leads to
\begin{equation}
\label{eq_point}
\fl
p_o=\Lambda-\frac{\xi^2}{3}-\frac{\sigma_o^2}{2},\quad \theta_o=-\xi,\quad \Pi_o=-\xi\,\sigma_o\cos\beta_o, \quad \mbox{and}\quad \sin\beta_o=0,
\end{equation}
with $p_o\doteq p(\theta,\sigma)|_{\theta=-\xi,\,\sigma=\sigma_o}$. From this, we see that, in an expanding universe scenario ($\theta\geq0$) without cosmological constant ($\Lambda=0$), the equilibrium points are possible only for fluids with negative pressure (in agreement with reference \cite{bnk}) and $\xi\cos\beta_o\le0$.

The case $\sigma(t)\rightarrow0$ implies $\Pi(t)\rightarrow0$ as expected, despite the fact that equations (\ref{sys_dyn_1})--(\ref{sys_dyn_4}) are not well defined for $\sigma(t)=0$ or $\Pi(t)=0$.
Equations (\ref{eq_point}) suggest that $\sigma_o=0$ implies $\Pi_o=0$. Let us illustrate the behaviour  of the system in the neighbours of $\sigma_o=0$ and $\Pi_o=0$. For the sake of simplicity, we also set $\xi=0$ and $H=H_0$ with $H_0$ a positive constant. A simple integration of equations (\ref{ee3}) and (\ref{therm_pi}) in this case yields, as we set $\pi_i(t_o)=\pi_{io}$ , $\sigma_{i}(t_o)=\sigma_{io}$ and $\kappa_o = 3H_0 - \frac{1}{\tau}$ ,
\begin{equation}
\frac{\pi_i}{\sigma_i}=\left\{
\begin{array}{ll}
\frac{\kappa_o \pi_{io}}{\pi_{io}+(\kappa_o\, \sigma_{io} - \pi_{io}){\rm e}^{-\kappa_o (t-t_o)}},&{\rm if}\quad \kappa_o \neq 0,\\[1ex]
\frac{\pi_{io}}{\sigma_{io}+\pi_{io}(t-t_o)},&{\rm if}\quad \kappa_o= 0.
\end{array}\right.
\end{equation}
Thus, for $\sigma_{io}=0$ and $\pi_{io}\neq0$, the ratio $\pi_i/\sigma_i$ is either a constant $3H_0-1/\tau$ or decreases as $(t-t_o)^{-1}$; for $\sigma_{io}\neq0$ and $\pi_{io}\neq0$, the ratio approaches zero with an exponential behaviour for $3H_0\tau<1$, or with a power law behaviour for $3H_0\tau=1$, or else approaches a non zero finite constant for $3H_0\tau>1$. The solution $\sigma_o=0$ and $\Pi_o=0$ is then a global attractor, which can be approached by either finite $\Pi/\sigma$ and $\sigma/\Pi$ for $3H_0\tau>1$ (and $3H_0\tau<1$ with $\sigma_{io}=0$) or else by $\Pi\ll\sigma$ for $3H_0\tau\leq1$ (with $\sigma_{io}\neq0$). In fact, it is not necessary to set $H(t)=H_0$ to obtain the qualitative global attractor behaviour. It is generally achieved by setting, in an expanding universe scenario, the energy condition $\rho+p\geq-\sigma^2$. This guarantees $\dot H\leq0$ from equations (\ref{ee1}) and (\ref{sys_dyn_3}), meaning that $H(t)$ is asymptotically approximated by its lower bound $H_0$.

The linearisation of equation (\ref{sys_dyn_5}) in the vicinities of the equilibrium points with $\beta=\beta_o$ yields
\begin{equation}
\label{sys_dyn_5_2}
\dot\beta=-(\cos\beta_o)(\tau^{-1}-\xi)\,(\beta-\beta_o).
\end{equation}

For $\xi<\tau^{-1}$, then $\beta\approx0$ fastly approach $\beta_o=0$ while $\beta\approx\pi$ fastly get away from $\beta_o=\pi$ (such behaviour is reversed for $\xi>\tau^{-1}$). This allows us to approximate the behaviour of the solutions in the vicinities of the equilibrium points by those with $\beta=\beta_o$ and with the help of equations (\ref{sys_dyn_3})--(\ref{sys_dyn_2}) only. This approximation has also the advantage of rendering the dynamical system autonomous. Below we discuss the case $\beta_o=0$.

The linear behaviour of equations (\ref{sys_dyn_3})-(\ref{sys_dyn_2}) applied at the condition (\ref{eq_point}) is given by
\begin{equation}
\label{lin_sys}
\left(\begin{array}{c}
\dot{\theta}\\[1ex]
\dot{\sigma}\\[1ex]
\dot{\Pi}
\end{array}\right)=
\left(\begin{array}{cccccc}
\xi-\fracc{3}{2}p_{\theta}&-\fracc{3}{2}\sigma_o-\fracc{3}{2}p_{\sigma}&0\\[1ex]
-\sigma_o&\xi&1\\[1ex]
0&-\fracc{\xi}{\tau}&-\fracc{1}{\tau}
\end{array}\right)
\left(\begin{array}{c}
{\theta-\theta_o}\\[1ex]
{\sigma-\sigma_o}\\[1ex]
{\Pi-\Pi_o}
\end{array}\right),
\end{equation}
where $p_{\sigma}\doteq \p p/\p \sigma$ and $p_{\theta}\doteq \p p/\p \theta$ with the derivatives evaluated at the equilibrium points $X_o$.
The characteristic polynomial for this system is
\begin{equation}
\label{ch_pol}
\begin{array}{l}
\fl
P(\lambda)\doteq \lambda^{3}+\left(\frac{3}{2}\,p_{\theta}- 2\,\xi +\frac{1}{\tau}\right)\lambda^2-\frac{1}{2} \left(3\,\sigma_o^2 + 3 \sigma_o \,p_{\sigma} - 2\xi^2 + 3\xi p_{\theta} + \frac{2\xi}{\tau} - \frac{3p_{\theta}}{\tau}\right)\lambda +\\[2ex]
\fl
\quad\qquad- \frac{3}{2} \frac{(\sigma_o + p_{\sigma}) \sigma_o}{\tau}=0.
\end{array}
\end{equation}
In general, the roots of $P(\lambda)$ are rather involved and do not give any physical insight about their properties. However, there are two special cases which we can analyse:\\

$(i)$ $\sigma_o\approx0$ (nearly isotropic equilibrium metric), the roots of $P(\lambda)$ are close to
\begin{equation}
\label{eigen_i}
\lambda_1=0,\quad \lambda_2=\xi-\frac{3}{2}p_{\theta},\quad \mbox{and}\quad \lambda_3=\xi-\frac{1}{\tau}.
\end{equation}
In virtue of the null eigenvalue, we cannot apply the Hartman-Grobman theorem and, therefore, the linear regime informs nothing about the solutions in the neighbors of such equilibria. Notwithstanding, if $p_{\theta}>2\xi/3$ and $\tau^{-1}>\xi$ at the equilibrium, then $\lambda_2<0$ and $\lambda_3<0$. Thus, the constant term in the characteristic polynomial determines the sign of $\lambda_1$ for $\sigma_o$ slightly greater than zero. Therefore, if $p_\sigma>-\sigma_o$ (respectively $p_\sigma<-\sigma_o$), then $\lambda_1>0$ ($\lambda_1<0$) and the equilibria are saddle points (respectively attractors) by continuity of the parameters. Furthermore, for $\xi\neq0$, equations (\ref{eq_point}) indicate that these equilibrium points behave as an effective cosmological constant since $\theta$ approaches a non-zero value while the anisotropies are washed out, leading to an isotropic configuration for the universe. The value of this effective cosmological constant is determined by the shear viscosity of the fluid. Note that equations (\ref{eigen_i}) hold as well for $\sigma_o=-p_{\sigma}$.

$(ii)$ For $\theta_o=0$, the roots of $P(\lambda)$ are
\begin{equation}
\label{eigen_ii}
\lambda_0=-\frac{1}{\tau},\quad \lambda_{\pm}= -\frac{3}{4}p_\theta \left[1\pm\sqrt{1 + 24\,\sigma_o\, \frac{p_\sigma + \sigma_o}{(p_\theta)^2}} \right].
\end{equation}
Note that two eigenvalues do not depend on the thermodynamical parameters $\tau$ and $\xi$. In this situation, $X_0$ may have all possibilities depending on the sign of the discriminant $\Delta\doteq (p_\theta)^2+24\,\sigma_o(p_\sigma+\sigma_o)$ and that of $p_\theta$. Roughly speaking, if $\Delta/(p_\theta)^2\leq1$ and $p_\theta>0$ ($p_\theta<0$), then $\lambda_{\pm}$ have negative (positive) real parts which means that the equilibria are attractors (saddle points); if $\Delta/(p_\theta)^2>1$, then the equilibria are necessarily saddle points for any $p_\theta\neq0$. Therefore, for a particular choice of the anisotropies, the Friedmann solutions with the asymptotic regime $\theta\rightarrow0$ when $t\rightarrow+\infty$ belong to a subset of equilibrium points of the anisotropic case.

\subsection{Barotropic fluids with linear equation of state}
Barotropic fluids are those whose pressure is a function of the energy density solely. Motivated by the current cosmological scenario, we assume the linear equation of state of a perfect fluid, namely $p=(\gamma-1)\rho$ with $\gamma\in[0,2]$ as a constant, allowing room for mechanical instabilities (for $0\leq\gamma<1$). Elimination of the energy density $\rho$ from the modified Friedmann equation (\ref{ee1}) in equation (\ref{sys_dyn_3}) yields
\begin{equation}
\dot{\theta}=-\frac{\gamma}{2}\,\theta^2-\frac{3(2-\gamma)}{4}\,\sigma^2 + \frac{3\gamma}{2}\,\Lambda.
\label{sys_dyn_5_prime}
\end{equation}
It is then easy to see that the equilibrium conditions require
\begin{equation}
\sigma_o^2 = \frac{2\gamma(3\Lambda-\xi^2)}{3(2-\gamma)}.
\end{equation}
The character of the eigenvalues of the system in the $\Lambda\neq0$ case is rather complicate. When $\Lambda=0$, it simplifies considerably and only the origin $X_0=0$ corresponds to an equilibrium point if $\xi=0$ and $0<\gamma<2$. The eigenvalues can be obtained from equation (\ref{eigen_i}) as $\lambda_{1},\lambda_{2}=0$ and $\lambda_{3}=-\frac{1}{\tau}$, and their corresponding eigenvectors are
\begin{equation}
v_{1}=(1,0,0),\quad v_{2}=(0,1,0),\quad v_{3}=(0,-\tau,1).
\end{equation}
When $\gamma=0$, the set of equilibrium points coincides with the $\theta$-axis and we are in a particular situation of case $(i)$. For $\gamma=2$, we obtain a set of equilibrium points given by the $\sigma$-axis and now the situation is a special case of $(ii)$.

\subsection{Fixed points at infinity}
We still assume here the linear equation of state previously used, and set the cosmological constant to zero for simplicity, in order to describe  the asymptotic behaviour of the system (\ref{sys_dyn_3})--(\ref{sys_dyn_4}) through the Poincar\'e compactification procedure \cite{llibre}. We will see that the asymptotic fixed points are sensitive to the critical values of the equation of state, namely $\gamma=0$ (cosmological constant) and $\gamma=2$ (stiff matter), for which the effective sound speed coincides in magnitude with the speed of light. Therefore, we  restrict our analysis to the cases $0\leq\gamma\leq2$, as well as, for reasons previously mentioned, to the set of equations (\ref{sys_dyn_3})--(\ref{sys_dyn_2}) with $\beta=0$ and all the three variables positive.

First of all, let us denote the points by $X=(x_1,x_2,x_3)\doteq(\theta,\sigma,\Pi)\in\mathbb{R}^3$. In practical terms, the compactification of $\mathbb{R}^3$ can be done by choosing $3$ local maps $\phi_k:\mathbb{R}^3\longrightarrow\mathbb{R}^3$ with $k=1,2,3$ defined as
\begin{equation}
\label{gen_map}
\phi_k(X)=\phi_k(x_1,x_2,x_3)\doteq\left(\frac{x_m}{x_k},\frac{x_n}{x_k},\frac{1}{x_k}\right),
\end{equation}
for $m<n$, with $m,n\neq k$, where $m,n,k=1,2,3$. For each chart, we shall denote $(z_1,z_2,z_3)\doteq\phi_k(x_1,x_2,x_3)$, such that the asymptotic behaviour of the dynamical system (\ref{sys_dyn_3})--(\ref{sys_dyn_2}) is provided by the evolution of the variables $z_i$'s and the fixed points at infinity, if they exist, lie always on the plane $z_3=0$. It should be noticed that the dynamics coming from each compactification procedure is not quantitatively equivalent to the original one due to a necessary time reparametrization. Each map $\phi_k$ determines a corresponding time stamp $\tilde t$ which is related to the time coordinate $t$ of the spacetime through $\tilde t=\int {\rm d}t/z_3(t)$. Thus, the prime here means derivative with respect to these different times $\tilde t$.

The asymptotic behaviour of equations (\ref{sys_dyn_3})-(\ref{sys_dyn_2}) according to $\phi_1$ is driven by
\numparts
\begin{eqnarray}
z'_1&=& -\frac{(2-\gamma)}{2}z_1 + \frac{3(2-\gamma)}{4}z_1^3 +z_2 z_3,\label{sys_dyn_phi3_1}\\[1ex]
z'_2&=&\frac{\gamma}{2}z_2 + \frac{3(2-\gamma)}{4} z_1^2 z_2 -\frac{z_2 z_3}{\tau} -\frac{\xi }{\tau} z_1 z_3,\label{sys_dyn_phi3_2}\\[1ex]
z'_3&=&\frac{\gamma}{2} z_3 + \frac{3(2-\gamma)}{4} z_1^2 z_3 ,\label{sys_dyn_phi3_3}
\end{eqnarray}
\endnumparts
where $(z_1,z_2,z_3)\doteq(\sigma/\theta,\Pi/\theta,1/\theta)$. Note that the three axes, besides being invariant under the above dynamics, contain all the equilibrium points. The equilibria which are relevant for the analysis of fixed points at infinity are divided in four disjoint sets:

\begin{enumerate}
\item{$(EP_{\phi_1})_1\doteq\left\{(z_1,z_2,z_3)\in\mathbb{R}^3\, | \, z_1=0,\, \forall z_2\neq0,\, z_3=0,\,\, \mbox{with}\,\, \gamma=0\right\}$. This is the case of a fluid dominated by a cosmological constant. Each of such equilibria  represents an asymptotic limit in a dynamics with $\theta$ diverging faster than $\sigma$ for arbitrary nonzero values of the ratio $\Pi/\theta$.}
\item{$(EP_{\phi_1})_2\doteq\left\{(z_1,z_2,z_3)\in\mathbb{R}^3\, | \, \forall z_1\neq0,\,z_2=0,\, z_3=0,\,\, \mbox{with}\,\, \gamma=2\right\}$. This is the case of a stiff matter fluid. Each of such equilibria  represents an asymptotic limit in a dynamics with $\theta$ diverging faster than $\Pi$ for arbitrary nonzero values of the ratio $\sigma/\theta$.}
\item{$(EP_{\phi_1})_3\doteq\left\{(z_1,z_2,z_3)\in\mathbb{R}^3\, | \, z_1=\sqrt{2/3},\,z_2=0,\, z_3=0,\, \, \mbox{with}\,\, 0\leq\gamma<2\right\}$. This equilibrium point represents an asymptotic limit in a dynamics with  $\theta$ going to infinity faster than $\Pi$, while keeping constant the relation $3 \sigma^2=2 \theta^2$. This situation corresponds to what is called viscoelastic energy in \cite{bnk}.}
\item{$(EP_{\phi_1})_4\doteq\left\{(z_1,z_2,z_3)\in\mathbb{R}^3\, | \, z_1=0,\, z_2=0,\, z_3=0,\,\, \mbox{with}\,\, 0\leq\gamma\leq2\right\}$. In this case, the equilibrium point represents an asymptotic limit in a dynamics with $\theta$ going to infinity faster than both $\Pi$ and $\sigma$.}
\end{enumerate}

The linearisation of (\ref{sys_dyn_phi3_1})-(\ref{sys_dyn_phi3_3}) furnishes that each set of equilibria leads to different eigenvalues, but the sub-space generated by the eigenvectors is the same, except for $(EP_{\phi_1})_4$. Namely, $(EP_{\phi_1})_1$ has a two-degenerate null eigenvalue associated with the eigenvector $v_{0}=(0,1,0)$ and $-1$ with eigenvector $v_{1}=(1,0,0)$. $(EP_{\phi_1})_2$ has a two-degenerate eigenvalue equal to $1$ with eigenvector $v_{0}$ and another one equal to zero with eigenvector $v_{1}$. $(EP_{\phi_1})_3$ has a two-degenerate eigenvalue equal to $1$ with eigenvector $v_{0}$ and another one equal to $2-\gamma$ with eigenvector $v_{1}$. This equilibrium point is a source. $(EP_{\phi_1})_4$ has a two-degenerate eigenvalue equal to $\gamma/2$ with eigenvectors $v_{0}$ and $v_{2}=(0,0,1)$ and another one equal to $-(2-\gamma)/2$ with eigenvector $v_{1}$. This equilibrium point is unstable and, for $0<\gamma<2$, it turns out to be a saddle point. Note that the plane $z_3=0$ is invariant under the action of the dynamics. Moreover, the solutions in the vicinities of this plane have $z_3'/z_3>0$, from which we conclude that all four equilibrium point sets are unstable; in particular, the two degenerate cases $(EP_{\phi_1})_1$ and $(EP_{\phi_1})_2$ are unstable.

The Poincar\'e map $\phi_2$ leads to the system
\numparts
\begin{eqnarray}
z'_1&=&-\frac{3}{4}(2-\gamma)+\frac{2-\gamma}{2}z_1^2-z_1 z_2 z_3,\label{sys_dyn_phi1_1}\\[1ex]
z'_2&=&z_1 z_2- z_2^2 z_3  - \frac{z_2 z_3}{\tau} - \frac{\xi z_3}{\tau},\label{sys_dyn_phi1_2}\\[1ex]
z'_3&=&z_1 z_3 -z_2 z_3^2,\label{sys_dyn_phi1_3}
\end{eqnarray}
\endnumparts
with $(z_1,z_2,z_3)\doteq(\theta/\sigma,\Pi/\sigma,1/\sigma)$. This system has four sets of relevant equilibrium points: \begin{enumerate}
\item{$(EP_{\phi_2})_1\doteq\{(z_1,z_2,z_3)\in\mathbb{R}^3\, | \, z_1=0,\,\forall z_2\neq0,\,z_3=0,\, \, \mbox{with} \,\, \gamma=2\}$. Each of such equilibria represents an asymptotic limit in a dynamics with $\sigma$ diverging faster than $\theta$ for arbitrary nonzero values of the ratio $\Pi/\sigma$.}
\item{$(EP_{\phi_2})_2\doteq\{(z_1,z_2,z_3)\in\mathbb{R}^3\, | \, z_1=0, \, z_2=0, \, z_3=0,\,\, \mbox{with} \,\, \gamma=2\}$. This equilibrium point represents an asymptotic limit in a dynamics with $\sigma$ diverging faster than both $\theta$ and $\Pi$.}
\item{$(EP_{\phi_2})_3\doteq\{(z_1,z_2,z_3)\in\mathbb{R}^3\, | \, \forall z_1\neq0,\,z_2=0,\, z_3=0,\, \, \mbox{with} \,\, \gamma=2\}$. This set is the same as $(EP_{\phi_1})_2$.}
\item{$(EP_{\phi_2})_4\doteq\left\{(z_1,z_2,z_3)\in\mathbb{R}^3\, | \, z_1=\sqrt{3/2},\,z_2=0,\, z_3=0,\, \, \mbox{with} \,\, 0\leq\gamma<2\right\}$. This equilibrium point is the same as  $(EP_{\phi_1})_3$}
\end{enumerate}
The linearisation of (\ref{sys_dyn_phi1_1})-(\ref{sys_dyn_phi1_3}) around these sets furnishes that $(EP_{\phi_2})_1$ has a three-degenerate null eigenvalue associated with the two eigenvectors $v_{0}=(0,1,0)$ and $v_2=(z_2+\xi+\tau z_2^2,0,\tau z_2)$. $(EP_{\phi_2})_2$ has also a three-degenerate null eigenvalue, but with two eigenvectors $v_{0}$ and $v_1=(1,0,0)$. $(EP_{\phi_2})_3$ has one null eigenvalue with eigenvector $v_{1}=(1,0,0)$ and a two degenerate eigenvalue equal to $z_1$ with eigenvector given by $v_{0}$. $(EP_{\phi_2})_4$ has a two-degenerate eigenvalue equal to $\sqrt{3/2}$ associated with only one eigenvector $v_0$ and another eigenvalue given by $(2-\gamma)\sqrt{3/2}$ with eigenvector $v_1$. Thus, \((EP_{\phi_2})_4\) is a source. Again, the plane $z_3=0$ is invariant under the action of the dynamics, but now the sign of $z_3'$ in the neighbours of this plane depends on the sign of $z_1-z_2z_3$. Thus, if \(z_1-z_2z_3>0\), as it is certainly the case for some among the relevant points which are close to \((EP_{\phi_2})_1\), as well as those which are close to \((EP_{\phi_2})_2\), we have instabilities. In the case $\gamma=2$, the plane $z_1=0$ is also invariant, and the solutions in the vicinities of this plane have $z_1'/z_1<0$, indicating that these equilibrium points have at least one attracting direction. Therefore, $(EP_{\phi_2})_1$ and $(EP_{\phi_2})_2$ are saddle points.

Finally, we get from $\phi_3$ the system
\numparts
\begin{eqnarray}
z'_1&=&- \frac{\gamma}{2}z_1^2-\frac{3(2-\gamma)}{4}z_2^2 +\frac{z_1 z_3}{\tau} + \frac{\xi z_1 z_2 z_3}{\tau},\label{sys_dyn_phi2_1}\\[1ex]
z'_2&=&z_3 - z_1 z_2 + \frac{z_2 z_3}{\tau} + \frac{\xi z_2^2 z_3}{\tau},\label{sys_dyn_phi2_2}\\[1ex]
z'_3&=&\frac{z_3^2}{\tau} + \frac{\xi z_2 z_3^2}{\tau},\label{sys_dyn_phi2_3}
\end{eqnarray}
\endnumparts
for $(z_1,z_2,z_3)\doteq(\theta/\Pi,\sigma/\Pi,1/\Pi)$. This system admits three sets of relevant equilibrium points:
\begin{enumerate}
\item{$(EP_{\phi_3})_1\doteq \left\{(z_1,z_2,z_3) \in \mathbb{R}^3\, | \, \forall z_1\neq0,\,z_2=0,\, z_3=0,\, \, \mbox{with} \,\, \gamma=0\right\}$. This case is the same as $(EP_{\phi_1})_1$.}
\item{$(EP_{\phi_3})_2 \doteq \left\{(z_1,z_2,z_3) \in \mathbb{R}^3\, | \, z_1=0,\, \forall z_2\neq0,\,z_3=0,\, \, \mbox{with}\,\, \gamma=2\right\}$. This set is the same as $(EP_{\phi_2})_1$.}
\item{$(EP_{\phi_3})_3\doteq \left\{(z_1,z_2,z_3)\in\mathbb{R}^3\, | \, z_1=0,\,z_2=0,\,z_3=0\,\, \mbox{with}\,\, 0\leq\gamma\leq2\right\}$. In this case, the equilibrium point represents an asymptotic limit in a dynamics such that $\Pi$ diverges faster than both $\theta$ and $\sigma$.}
\end{enumerate}
The linearisation of the above equations around the sets of equilibria furnishes that the set $(EP_{\phi_3})_1$ has a two-degenerate null eigenvalue associated with the eigenvector $v_1=(1,0,0)$ and another one given by $-z_1$ with eigenvector $v_0=(0,1,0)$. The sets $(EP_{\phi_3})_2$ and $(EP_{\phi_3})_3$ have three-degenerate null eigenvalues, but the former has the eigenvectors $v_2=(z_2+\tau+\xi z_2^2,0,\tau z_2)$ and $v_0$ while the latter possesses the eigenvectors $v_1$ and $v_0$. Since $(EP_{\phi_3})_1$ and $(EP_{\phi_3})_2$ are equivalent to sets of equilibrium points previously discussed, it remains to analyse $(EP_{\phi_3})_3$ solely. Once again, the plane $z_3=0$ is invariant and, for \((1+\xi z_2)/\tau>0\) (as it is the case for the relevant points which are close to both this plane and \((EP_{\phi_3})_3\)), we have $z_3'/z_3>0$. Therefore, $(EP_{\phi_3})_3$ is unstable.

As a result of the above three compactification procedures, we obtained all possible cases of divergence for any of the variables \(\theta,\,\sigma,\,\Pi\). All such cases are unstable, with behaviours similar to either source or saddle points. Part of these instabilities could be forecasted from a simple analysis of equation (\ref{lean_rel}) and its implications on the evolution of the anisotropies. Therefore, it is expected that the solutions which dynamically evolve to such divergent configurations should be very special ones, in the sense that a small deviation from it yields a totally different global behaviour (even though this may not be true for {\em all} possible small deviations).

\section{Concluding remarks}
It is well-known that the homogeneity and isotropy of the Universe cannot be directly measured \cite{roy,wiltshire} at all scales due to our condition as local observers. Nowadays, it has been suggested some gravitational tests of the {\it cosmological principle} (homogeneity plus isotropy) \cite{roy,wiltshire,saadeh}. This special role played by homogeneity makes the investigation of Bianchi models particularly relevant. As we have seen, they favour the isotropic models for late times when the viscous fluids evolve according with the thermodynamic laws towards the isotropic equilibrium in most of the cases.

Furthermore, with the aid of the standard deviation variables, the exact solutions we found here show that Bianchi-I models allow for a relic cosmic acceleration in the future if the possible anisotropies of the past have a thermodynamical origin and their evolution was driven by the Israel-Stewart equations. Although the anisotropies in the space-time might be very small on account of present observations, a concrete cosmological model from Bianchi-I spacetimes is still under investigation \cite{saadeh}. Our results indicate that an alternative explanation of the exotic matter components (the so-called {\em dark energy}) could be provided in terms of matter anisotropies only.

Finally, we have tried here to study a simple case of barotropic fluids with the anisotropic pressure without any {\it ad hoc} restriction on the phase space. We reduced the original highly non-linear dynamical system in $\mathbb{R}^6$ to another one in terms of a second order degree polynomial in $\mathbb{R}^3$ by taking into account physical arguments solely. We have then seen that the dynamics in the vicinities of the singular points is very slow and degenerate, admitting several null eigenvalues, most of which are unstable. This preliminary result presages that a comprehensive analysis of this situation, which is still an open question, may unhide interesting and non-trivial features, as suggested by the outcomes of the BKL conjecture.

\section*{Acknowledgments}
We would like to thank Lucas R. dos Santos for valuable comments on the qualitative analysis. Part of this work was done when EB was financially supported by the PNPD/CAPES program (grant No. 1614969).

\appendix
\section{Diagonalizability of Bianchi-type I geometries}
It is possibly known that not all Bianchi-type I geometries do admit a diagonal form. Indeed, rigorous treatments of Bianchi-I metrics do not assume a diagonal representation (see \cite{Misner}, for instance). However, this is apparently not agreed by all of the literature (see theorem 4.2 in \cite{elismac} and equation 18.3 in \cite{elismartmac}, for instance). In order to definitely clarify this point, let us prove in the following that the Bianchi-type I geometry determined by the line element
\begin{equation}
\label{ds2us}
{\rm d}\tilde s{}^2=-{\rm d}\tilde t{}^2+{\rm d}\tilde x{}^2+\tilde t{}^2{\rm d}\tilde y{}^2+\frac{1}{\tilde t{}^2}\,{\rm d}\tilde z{}^2+2\tilde t\,{\rm d}\tilde x\,{\rm d}\tilde y
\end{equation}
cannot be cast in the diagonal form of equation (\ref{ds2-bian_l}) by any coordinate transformation.\footnote{We are not interested in any physical interpretation of the geometry in  equation (\ref{ds2us}), thus we are not claiming that such an interpretation exists. The form of \(\tilde g{}_{\mu\nu}\) was chosen on mathematical grounds only.}
This is accomplished by the fact (to be shown below) that the assumption of the existence of such a coordinate transformation necessarily leads to a contradiction. Thus, let us temporarily assume that such a coordinate transformation exists.

The Bianchi-type I diagonal geometry in equation (\ref{ds2-bian_l}) is given in a form which explicitly admits a vector basis \((e_0,\,e_1,\,e_2,\,e_3)\) which is integrable \([e_\mu,\,e_\nu]=0\) and such that \(e_i\) are three spacelike Killing vectors \({\cal L}_{e_i}g=0\), where the four vectors \(e_\mu\) are defined as
\(e_0=\partial_t:=\partial/\partial t,\,e_1=\partial_x:=\partial/\partial x,\,e_2=\partial_y:=\partial/\partial y,\,e_3=\partial_z:=\partial/\partial z. \)
Here \({\cal L}_X\) denotes the Lie derivative with respect to the vector \(X\), and \({\cal L}_XY=[X,\,Y]\) for arbitrary vectors \(X,\,Y\). Moreover, we will make use of the identities \({\cal L}_{\alpha X+\beta Y}\,f=\alpha\,{\cal L}_X\,f+\beta{\cal L}_Y\,f\) and
\[{\cal L}_Z\{g(X,\,Y)\}=({\cal L}_Z\;g)(X,\,Y)+g({\cal L}_ZX,\,Y)+g(X,\,{\cal L}_ZY)\]
which hold for arbitrary function \(f\), vectors \(X,\,Y,\,Z\) and scalars \(\alpha,\,\beta\) if \(g\) is the metric tensor.

First of all, let us consider \(t=t(\tilde t,\,\tilde{\vec x})\). If \(\tilde\partial{}_i\,t=\partial t/\partial\tilde x{}^i\neq0\), then the homogeneity surfaces \(\Sigma\) of equation (\ref{ds2-bian_l}) does not coincide with the homogeneity surfaces \(\tilde\Sigma\) of equation (\ref{ds2us}). Namely, we would have \(\tilde\Sigma{}_{\tilde t=\tilde k}\neq\Sigma_{t=k}\) for any possible choices of \(k,\tilde k\in\mathbb{R}\). Therefore, the two foliations of the spacetime by such homogeneity surfaces are such that the two classes of surfaces intercept each other. The homogeneity orbit of any given point turns out to be the whole spacetime, thus implying that the spacetime is homogeneous. This, in turn, requires that the directional scale factors \(l_1(t),\,l_2(t),\,l_3(t)\) are all independent of time \(t\), from which the spacetime is locally flat. But the Ricci curvature scalar for the metric in equation (\ref{ds2us}) is \(R=5/(2\tilde{t}\,{}^2)\), which is the expected contradiction. Therefore, it follows that \(t=t(\tilde t)\) only.
From this, it yields the two-fold result \({\rm d}\tilde t={\rm d}t/a(t)\) (and \(\tilde{\partial}{}_{t}:=\partial/\partial\tilde t =a(t)\partial_t\)) with \(a(t)=1/({\rm d}\tilde t/{\rm d}t)\), and \(\tilde{\partial}{}_i:=\partial/\partial{\tilde x{}^i}=a_i^j\,\partial_j\) with $a_i^j=a_i^j(t,\,\vec x)$.

Then, let us explore the integrability condition which defines  the coordinate basis \((\tilde e{}_0,\,\tilde e{}_1,\,\tilde e{}_2,\,\tilde e{}_3)\) for which that equation (\ref{ds2us}) takes a diagonal form, with \(\tilde e{}_i\) being spacelike Killing vectors. From the above reasoning, we have \(\tilde e{}_0=a\,e_0\) and \(\tilde e{}_i=a_i^j\,e_j\). We should have
\[\hspace{-3em}
0=[\tilde e{}_0,\,\tilde e{}_i]=[a\,e_0,\,a_i^j\,e_j]=a\,a_i^j[e_0,\,e_j]-a_i^j(\partial_j\,a)e_0+a(\partial_t\,a_i^j)e_j
=a(\partial_t\,a_i^j)e_j
\]
where use has been made of \(\partial_j\,a=0\). It follows that \(\partial_t\,a_i^j=0\), thus yielding \(a_i^j=a_i^j(\vec x)\).

Now we turn our attention to the Killing condition of \(\tilde e{}_i\), which means \({\cal L}_{\tilde e{}_i}g=0\). We can most conveniently evaluate this equation in the basis \((e_\mu)\) as \(g_{\mu j}a_i^j{}_{,\nu}+g_{\nu j}a_i^j{}_{,\mu}=0\). Explicit calculation of these Killing conditions for the metric in equation (\ref{ds2us}) yields \(a_i^j{}_{,k}=0\). Thus, we obtain \(a_i^j{}_{,\nu}=0\), from which it follows that all \(a_i^j\) are constant. The coordinate transformation can then be read in the differentials of the coordinates in the form
\begin{equation}
\label{change}
\pmatrix{{\rm d}\tilde t\cr{\rm d}\tilde x\cr{\rm d}\tilde y\cr{\rm d}\tilde z}
=\pmatrix{1/a(t)&0&0&0\cr 0&A&B&C\cr 0&D&E&F\cr 0&G&H&J}
\pmatrix{{\rm d}t\cr{\rm d}x\cr{\rm d}y\cr{\rm d}z}
\end{equation}
for convenient constants \(A,\,B,\,C,\,D,\,E,\,F,\,G,\,H,\,J\). Direct substitution of equation (\ref{change}) into equation (\ref{ds2us}) immediately gives the line element in terms of the $(t,\,x,\,y,\,z)$ variables. A straightforward comparison with equation (\ref{ds2-bian_l}) then leads to \([a(t)]^2=1\) (from which \(a(t)=1\) in order to preserve the future orientation) together with
\begin{eqnarray}
\left\{\begin{array}{l}
[l_1(t)]^2=A^2+2ADt+2D^2t^2+G^2/t^2\cr
[l_2(t)]^2=B^2+2BEt+2E^2t^2+H^2/t^2\cr
[l_3(t)]^2=C^2+2CFt+2F^2t^2+J^2/t^2
\end{array}\right.
\label{eles}\\[1ex]
\left\{\begin{array}{l}
AB+(AE+BD)t+2DEt^2+GH/t^2=0\cr
AC+(AF+CD)t+2DFt^2+GJ/t^2=0\cr
BC+(BF+CE)t+2EFt^2+HJ/t^2=0
\end{array}\right.
\label{constraints}
\end{eqnarray}
Elementary algebraic manipulations yield that all the involved products of pairs of constants in equation (\ref{constraints}) are zero. A little bit more of algebra then yields that all solutions of equation (\ref{constraints}) are such that the directional scale factors given by equation (\ref{eles}) do satisfy the equation \(l_1(t)\,l_2(t)\,l_3(t)=0\). But this would render equation (\ref{ds2-bian_l}) into a singular metric, which is obviously unacceptable. This is the expected contradiction foreseen at the beginning of the appendix, from which we definitely conclude that equation (\ref{ds2us}) indeed cannot be cast in the diagonal form of equation (\ref{ds2-bian_l}). Therefore, not all the Bianchi-type I geometries are always diagonalizable, and the assumption of a diagonal form for the Bianchi-I metric (as it was done in the body of this work) actually restricts the class of Bianchi-type I geometries to a proper subset of it.

\end{document}